# Prediction of Domain Values: High throughput screening of domain names using Support Vector Machines

Zsolt Bikadi[1], Sapumal Ahangama[2], Eszter Hazai[1,*]
1 Virtua Drug Ltd, 4C Csalogany, Budapest, Hungary
2 National University of Singapore, School of Computing, 13 Computing Dr, Singapore 117417

*corresponding author

## Abstract

As connected devices multiply and the internet matures into a ubiquitous platform for exchange and communication, the question of what makes a domain name valuable is ever more significant. Due to the scarcity of meaningful vocabulary and the persistence of domain-related data, the buying and selling of previously owned domain names, also known as the domain aftermarket, has evolved into a billion dollar industry. Each day over a 100,000 domain names expire and become available for re-registration. Manual appraisal is impossible at such a volume; thus a method for the automated identification of valuable domain names is called for. The aim of our study was to develop a method for high throughput screening of domain names for rapid identification of the valuable ones. Five different aspects that make a domain name valuable were identified: name quality, domain authority, domain traffic, active domain age and domain health. An SVM method was developed for high throughput screening of domain names. Our method was able to identify valuable domain names with 97% accuracy for the test set and 93% for the external set and can be used for routinely screening the domain aftermarket.

## Introduction

Getting the whole world online is a goal considered worthy by many, from corporations like Google and Facebook to the UN. Being online means being able to visit any internet-accessible domain and being able to create your own; as such, the World Wide Web is a made up of these active namespaces. Over the past two decades innumerable spaces have come and gone, while some innovative domains have become synonymous with internet-enabled ways of doing everyday things like searching for information or booking a taxi. In a mobile era where the term "Web 2.0" seems woefully dated, it has become a truism that all the "good" domain names are "already taken".
In its simplest form, a domain name consists of a name string (e.g., "Google") and an extension (e.g., ".com"), also known as a Top-Level Domain (TLD). In order to function as an address on the web, this name must be unique. The Internet Corporation for Assigned Names and Numbers (ICANN) coordinates domain registrar databases across the world containing over 290 million registered domain names and over a thousand TLDs, with .com being the most popular (https://www.verisign.com/assets/domain-name-report-december2015.pdf).

Names are principally registered on a first-come, first-served basis, although several new extensions offer priority registration for trademark holders. Domain owners can fully use the domain and maintain or transfer its ownership as long as they bear the moderate fixed recurring costs levied by the registrar (Burshtein 2005).

Domain ownership can be lost due to non-payment of the registrar's renewal fees, and also if there is some bad faith or lack of legitimate use on the part of its owner. An active secondary market encompassing both owned and expiring domains has evolved for buying and selling domain names. Most valuable domains are often bought for thousands of dollars. According to dnjournal.com, 85 domain names were sold for over 100,000 USD in 2015, with the highest sales being in the million dollar range.

It is important here to make a distinction between domain names and websites, although obviously they are not independent. Websites are collections of related web pages served from the same domain or subdomain and as such incur additional costs of hosting, maintenance and development. In the context of sales, a domain-only sale is different from one which includes an associated website or business being transferred to the new owner. The frame of this study focuses on domain-only sales and valuations.

Domaining is the practice of identifying and acquiring domain names as an investment with the intent of making profit. Typically, profits can be made by brokering, developing a website, using the domain for search engine optimization or domain parking. The domain space is infinite (i.e. the number of domain names is infinite), while the number of valuable domains is limited. Thus, effort has been made over the years in order to effectively valuate domains. While domain names are valuable assets, they can be illiquid, especially when they are for sale at a fixed price. Since domain owners can set a fixed price independently of the market value, fixed price transactions and private transactions do not necessarily reflect the true market value of a domain name. However, the existence of open secondary markets in the form of domain auctions without a reserve price assures a fair market valuation for such domains. Bidders determine the value of a domain based on many parameters, such as meaning, length of the domain name, previous content, previous popularity and links pointing to the domain name. The highest bid is the sales price of the domain and the highest bidder becomes the new owner of the domain. While obviously there are subjective factors when buyers valuate a domain, sales prices are not randomly distributed but are dependent on different domain parameters.

The domain name market as a whole has been analysed by Lindenthal (Lindenthal 2014), (Lindenthal 2016) using the database of Sedo.com. A constant quality price index was developed to illustrate the parallels between domain market and stock market behaviours. In his study differences in the quality of individual domains were only considered in terms of different domain extensions. Our study is different in this view: the quality of individual domain names was explored and predicted. Most importantly, as our study shows, domain name value derives not only from its name but from its history as well, and both these characteristics can be well quantified.

A domain name can arrive in the aftermarket in one of two ways: either the owner of the domain wishes to sell it or the domain name is not renewed and thus the registrar or its partner sends the domain to an auction. Buyers wishing to acquire currently registered domains can also "backorder" them through various companies, in which case the registrar attempts to acquire the domain on behalf of the potential buyer in the event that it becomes available again.

About 27% of registered .com and .net domain names expire every year, and this ratio has been stable through the last several years (see Verisign's industry report). Consequently, more than 100,000 domain names become available on these aftermarkets every day. Thus, it can be inferred that many valuable domain names are not sold through the secondary market, but rather become available for registration after expiring. Plenty of websites offer direct access to expired domains via auction, such as NameJet, Snapnames, Sedo, Dropcatch and GoDaddy, with the latter possessing the highest number of domain names for sale.

As the typical backorder price of a domain is close to 100 USD, a valuable domain may be defined as a domain worth more than 100 USD. A method which is able to analyse thousands of domain names on a daily basis in order to preselect valuable ones that might be worth backordering would be of eminent use. Various domain valuation services are available online, with Estibot being one of the oldest and most well-known. However, their estimation methodology has never been published and cannot be consistently inferred from the valuation results. Furthermore, new, albeit scattered services have emerged to scrutinize various aspects of domain names. A comprehensive methodology that employs up-to-date techniques, investigates the significance and limitations of available data and is open to discussion is thus called for.

In the first stage of this study, factors affecting domain value were identified on the basis of which a high throughput method was developed for domain screening. The principal concern of the method was the ability to automatically screen hundreds of thousands of domains a day and classify them into valuable and non-valuable ones with high accuracy. In consonance with the "no free lunch" theorem there is no classifier that would perform best on all possible problems; however, SVM classifiers routinely rank among the best for a wide range of problems when evaluated over many real-life data sets (van Gestel et al. 2004). Our research group has successfully utilized SVM in bioinformatics for predicting albumin binding sites, BCRP and P-gp substrate properties (Bikadi et al. 2011), (Zsila et al. 2011), (Hazai et al. 2013)A similar methodology can be applied for high-throughput screening of domain names as well.

The support vector machine (SVM) method is a machine learning technique introduced by Vapnik et al. (Vapnik 1995) that has been successfully used in a wide range of applications ranging from linguistics to financial forecasting, image classification and medicine (Mjolsness 2001), (Noble 2006), (Tian, Shi, and Liu 2012), Hazai 2013. The SVM model treats the entities that are to be classified as points in a high-dimensional space defined by distinctive features, or so-called "descriptors", in order to find a hyper-plane that separates them. The selection of this hyper-plane maximizes the capability of the SVM to predict the correct classification of new entities. Various nonlinear functions called kernels can be used to find the best separating surface. The most widely used kernels are polynomials, Gaussian functions or RBF kernels (Cristianini and Shawe-Taylor, n.d.)

In the frame of our study, an SVM method was developed for the separation of valuable domains from non-valuable ones. This method is founded on and in turn tests our premise that domain value is influenced by a combination of quantifiable parameters and can be predicted using a well-built mathematical model.

## Methods

### Data set

In order to prepare sets for the classifier, domain name auction prices were collected from three sources:
i. making an initial bid on GoDaddy, NameJet and Snapnames auctions to secure access to the final sales prices;
ii. collecting recent data of the same marketplace sales from Namebio, a database of domain name sales;
iii. accessing GoDaddy closeouts domains, comprising domains that were not sold during a 7 day auction, via expireddomains.net.
Only domain auctions without reserve price with at least 2 concurrent bidders for the same domain name were included. Altogether, closing prices and sales data for 903 domain name sales were collected between March, 2015 and February, 2016. At the time of each auction the following properties were collected for the associated domain name: Google PageRank; Alexa rank (alexa.com); SimilarWeb rank (similarweb.com); Moz domain authority (opensiteexplorer.com); Moz page authority for the www subdomain; Moz number of backlinks; SEOkicks domainpop (seokicks.de); number of snapshots in Wayback Machine (archive.org); date of the first snapshot in Wayback Machine; domain availability with the most popular domain extensions (.com, .net, .org, .info, .biz and .us); google.com search volume of the domain keywords (using semrush.com); the number of URLs indexed in Google; Google AdSense blocked status and park blocked status. Domain Value was used to query and export all these parameters with the exception of SimilarWeb, Google AdSense blocked status and Google indexed status which were collected manually on similarweb.com, ctrlq.com/sandbox and using site:domain_name search on google.com, respectively.

### Hierarchical clustering and Support Vector Machine (SVM) calculations

The 903 domain names were grouped into equal sized training, test and external data sets which were carefully chosen so that all three data sets covered the whole "domain space"; the diversity of the training set was maximized. Since the number of domain parameters (potential descriptors) are limited, instead of feature selection all available parameters were used to build an appropriate SVM model. Feature extraction was carried out by hierarchical clustering of domain parameters using Cluster 3.0 software (de Hoon et al. 2004). Spearman's rank correlation was used for hierarchical clustering as it is less affected by the presence of outliers than Pearson correlation coefficient. Domain properties in the same cluster were combined into a complex descriptor using their geometric mean. Based on the correlation of properties, ACR was used in both authority and traffic descriptor as it was shown to possess a high correlation with both authority and traffic properties. DOB was used for name description as well as age because of the same reason. The Libsvm software (Chang and Lin 2011) was used for SVM calculations using five complex descriptors. Linear, polynomial, and radial basis function (RBF) kernels were tested. Gaussian RBF kernel showed higher prediction accuracy in all investigated cases, therefore a Gaussian RBF was chosen as the kernel function in our SVM calculations. In the training process, the regularization parameter (C) and the kernel width parameter (γ) were optimized using a grid search approach. The best combination of C and γ was selected by a grid-search with exponentially growing sequences of C and γ. Prediction power of the SVM model was evaluated based on the number of true positive (TP), true negative (TN), false positive (FP), and false negative (FN) predictions. Accuracy (ACC), sensitivity (SE), specificity (SP) and Matthews's correlation coefficient (MCC) were calculated using the equations given below (Baldi et al. 2000).

$$ACC = (TP+TN)/(TP+TN+FP+FN)$$
$$SE = TP/(TP+FN)$$
$$SP = TN/(FP+TN)$$
$$MCC = (TP \times TN - FP \times FN)/\sqrt{(TP+FN)(TP+FP)(TN+FP)(TN+FN)}$$

Accuracy is a measure of true hits in the entire calculation with both valuable and non-valuable domains included. Sensitivity (true positive rate) reflects the prediction accuracy of valuable domains, while specificity (true negative rate) is the measure of the prediction accuracy of non-valuable domains. The Matthews correlation coefficient considers over- and under-prediction and provides a more balanced evaluation of prediction than accuracy. MCC = 1 means a perfect prediction, MCC = -1 is a total negative correlation, whereas MCC = 0 indicates a random prediction. The prediction power of our model was tested on an external data set that was not part of the model building procedure.

## Results and Discussion

In the first part of our study the parameters influencing domain auction values were defined. There are publications indicating the commercial significance of words and their meanings in the context of domain names (Lindenthal 2014), (Lindenthal 2016). Effective marketing in search engine results depends at least in part on domain names *(Mueller 1998)*. It has been shown that in about a quarter of all cases the domain name itself influences the click through rate of a search result (Ieong et al. 2012)*.* However, in Google Search's ranking algorithm the number and quality of external links (backlinks) pointing to a domain name are even more important than the words or keywords in the domain name (Brin and Page 2012). Without doubt, the success of most websites depends heavily on the ability of internet users to find it quickly, primarily via Google search. Indeed, a whole industry named Search Engine Optimization (SEO) has emerged (Evans 2007) to finesse the ranking and appearance of websites in the top results for desirable keywords. In accordance with the importance of backlinks, our study shows that a catchy or meaningful name is just one factor of the domain auction price; the history of the domain and what was hosted on it also affects its current value. If a site was previously hosted under a domain and was linked to from other sites on the web, the domain name can continue to derive positive (or negative, in case of low quality backlinks) value from the extant backlinks, even after the site itself has been terminated.

Various online tools are available to examine data relevant to the name and history of a domain. One approach to determine the potential relevance or commercial value of a domain name is to examine the search volume of the words in the domain name (e.g., in the case of englishlanguage.com, the search volume of the phrase "English Language" would provide a clue to the volume of users searching for content related to these keywords). Investigating domain history includes exploring the number and quality of existing links to the domain and estimating its current and past traffic and popularity.

Due to the web's inherent decentralization, comprehensive data regarding sites is not collected in any one place. Over 30 trillion web pages are currently indexed by Google for search purposes, yet this accounts for less than 20% of the whole web according to estimates (Lawrence and Giles 2000), and no public-facing tool offers comprehensive information regarding the network of backlinks in all Google indexed sites. The quantified measure of a domain's backlink profile can therefore not be fully accurate. Similarly, current and past traffic is not publicly available. Furthermore, the open and increasingly profitable character of the internet exposes nearly all parameters to a certain degree of manipulation, and this manipulation has to be filtered out in order to correctly determine domain strength. Thus, although some parameters overlap in terms of their investigated source and impact (e.g., backlink profile or traffic), the determination of the real value of a domain can only be carried out by considering multiple available domain parameters and their correlations. The investigated tools all yield quantitative measures of domains and are listed and discussed below. The tools were grouped according to their investigated aspects: backlink profile, history, traffic, age, name and domain health.

| Analysed aspect | Domain descriptor | Abbreviation | Link |
|---|---|---|---|
| Backlink profile | Google Pagerank | **PR** | http://www.google.com/toolbar |
| | MOZ parameters Domain Authority. Page Authority. Backlinks | **DA. PA. BL** | http://moz.com/researchtools/ose/ |
| | SeoKicks DomainPop. LinkPop | **DP** | http://en.seokicks.de |
| History | Archive Snapshot Count | **ACR** | http://archive.org/web |
| Traffic | Alexa rank | **Alexa** | http://www.alexa.com |
| | SimilarWeb rank | **Similarweb** | http://www.similarweb.com |
| Age | Registration year | **DOB** | http://archive.org/web |
| **Name** | SEMrush Search Volume Phrase | **SV** | http://www.semrush.com |
| | Total Extension count | **TE** | http://whois.domaintools.com |
| | Properties of the domain name: length. hyphen and number | **LE. HY. NU** | |
| **Health** | Deindexed | **SB** | http://google.com. site:example.com |
| | Parking blocked | **PB** | https://www.bodis.com/account#domain-status-tool |
| | Adsense blocked | **AB** | http://ctrlq.org/sandbox |

Table 1: Tools for characterizing domain quality

## Domain Backlink Profile Properties

**Google Pagerank (PR)**: From the beginning of Google, its founders considered the ranking of results as a vital aspect of the search function (Brin and Page 1998). Current Google search rankings are famously complicated, occasionally unpredictable, and said to rely on "over 200 unique signals or clues" regarding each site. Among these, PageRank is the most well-known, and until recently, was the one publically released 'clue' by Google regarding the value of a site's backlink profile. The original concept of PageRank was published early in Google's evolution and since then has received more than 14,000 citations (Brin and Page 1998). While its exact formula has never been revealed, the calculation of a domain's PageRank is based on the principle that the quality and relevance of a web page can be deduced to a significant extent by the number, quality and content of other websites that link to it. Like Google's search algorithm itself, PageRank is continuously recalculated and remains a significant part of Google's ranking algorithm. Its value for a page used to be publically available as a logarithmic scale between 0 and 10. However, due to the importance of PageRank and the appearance of search engine 'gamers' or 'black hatters' (referring to those that actively employ duplicitous means to trick Google into valuating their pages higher), PageRank is no longer publically updated as of 6[th] December 2013. Thus the PageRank of a domain reflects its "frozen state" as observed in 2013. In March 2016, Google has published discontinuing the PageRank toolbar. Thus, the "frozen state" PRs are even harder to access.

Several algorithms have been suggested for the simulation of Google's PageRank calculation (Ishii and Tempo 2010), (Gleich et al. 2010). Furthermore, the underlying concept of the original PageRank has been utilized in various fields such as analysing protein interactions (Ivan and Grolmusz 2010), measuring author impact (Yan and Ding 2011) and opinion formation of social networks (Kandiah and Shepelyansky 2012).

As search engines are not transparent regarding their ranking methods, their algorithm can only be inferred from their search results pages and by exploration of their patents. In this vein, several tools have been made to evaluate domains from various perspectives, with MajesticSEO, ahrefs, Quantcast, SimilarWeb, Moz and SEOkicks are the most popular ones. In our work, MOZ and SEOkicks parameters were utilized as they provided free full access to the data, thus enabling the processing of vast amounts of data.

**MOZ Domain Authority (DA), Page Authority (PA) and MOZ backlinks (BL)**: Moz.com's backlinks service calculates the number of backlinks pointing to a given URL (Mavridis and Symeonidis 2014). A search engine optimization hub, Moz collects backlink data for millions of web pages and makes it available through an API. 'Authority', which conceptually mimics PageRank, is scored between 0 and 100 for both the domain itself (Domain Authority) and any public-facing pages on the site (Page Authority), with a higher number indicating more authority and more likelihood of appearing in top of search results. Using more than 40 signals, Moz employs a machine learning method for calculating DA and PA.

**SEOkicks Domain Pop (DP)**: SEOkicks is another service that maintains and provides access to backlink database for domains. The score is not logarithmic and does not have upper limits. DP stands for Domain Popularity i.e. the number of backlinks from different domains (Alpar, Koczy, and Metzen 2015).

Since the parameters described above are exploring the same domain property they are of course expected to correlate. However, as each service maintains its own databases (with Google possessing the highest amount of raw data and processing power) and none of them provide access to the data itself, there is difference in their valuation of the same domain, as well as value in incorporating multiple indicators, particularly in lieu of public PageRank updates. **Figure 1** shows the correlation of PR with DA, which claims to simulate PR value. As can be seen, the correlation is not high enough to justify choosing a single parameter to encapsulate and quantify a domain's backlink profile.

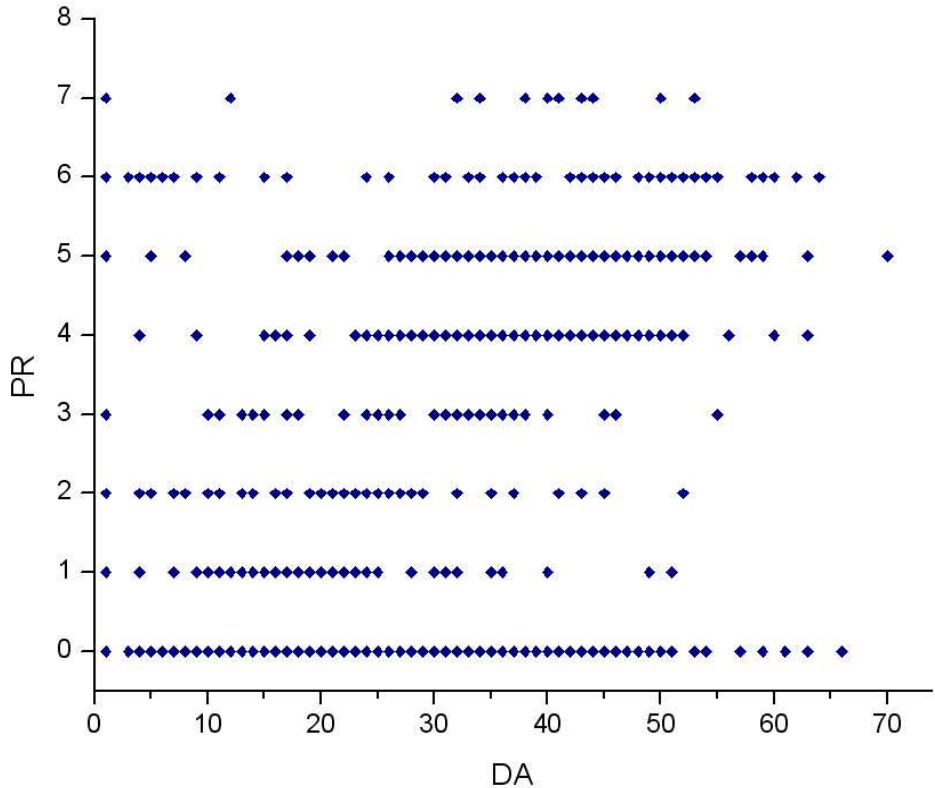

**Figure 1:** Relationship between PR and DA of the 903 investigated domains

## Domain History Properties

**Archive Number of Snapshots (ACR) and Archive Year (ABY)**: Archive.org's Wayback Machine is the web's largest and oldest archiving service, containing over 470 billion web captures as of March, 2016, dating all the way back to 1996 (Murphy, Hashim, and O'Connor 2007), (Kahle 1997). By searching for a URL users can see any and all snapshots taken by Archive.org of the web pages on that domain; in this way, Wayback Machine is the primary way that even insignificant data stays accessible on the web long after it has been removed, either by the domain owner or due to domain expiration. While Archive.org has never publically published how it determines when to capture a page, more popular sites are usually captured more frequently than others. Thus the number of archive.org snapshots can signal a domain's popularity, if not for the fact that snapshots can be manually triggered. However, ABY, which reflects the oldest date a domain was captured, is significant because it can offer a more accurate signal of the 'public' age of a domain as compared to the date of domain registration. A domain can be registered but never turned into a site, thus lacking a meaningful history or backlink profile. An expired domain also has its registration date overwritten, making it an unreliable factor when considered in isolation.

## Domain Traffic Properties

**Alexa rank**: Alexa rank is a widely used and well regarded metric for measuring website or domain traffic (Olteanu et al. 2013). Domains are ordered according to the average number of daily visitors to the site and the number of page views it has received over the previous three months. Alexa rank #1 is the highest rank, signifying the most visited website in the world. However, Alexa's methodology leaves something to be desired as it is extrapolated from a small and self-selecting set of users who have installed the Alexa toolbar. Thus, at low traffic the estimation is unreliable.
**Similarweb Rank:** SimilarWeb also aims at traffic estimation. They claim to use various channels for data collection such as monitoring more than 200 million devices, getting data from internet service providers, direct measurements from directly connected apps and websites etc. The ranking system is similar to that of Alexa; SimilarWeb's #1 is their estimation of the most visited website in the world.
These measures analyse the same aspects, i.e. traffic of a domain, however, they show rather poor correlation especially at low traffic (higher number rank) as presented in **Figure 2**. This low correlation indicates that traffic prediction is of low accuracy.

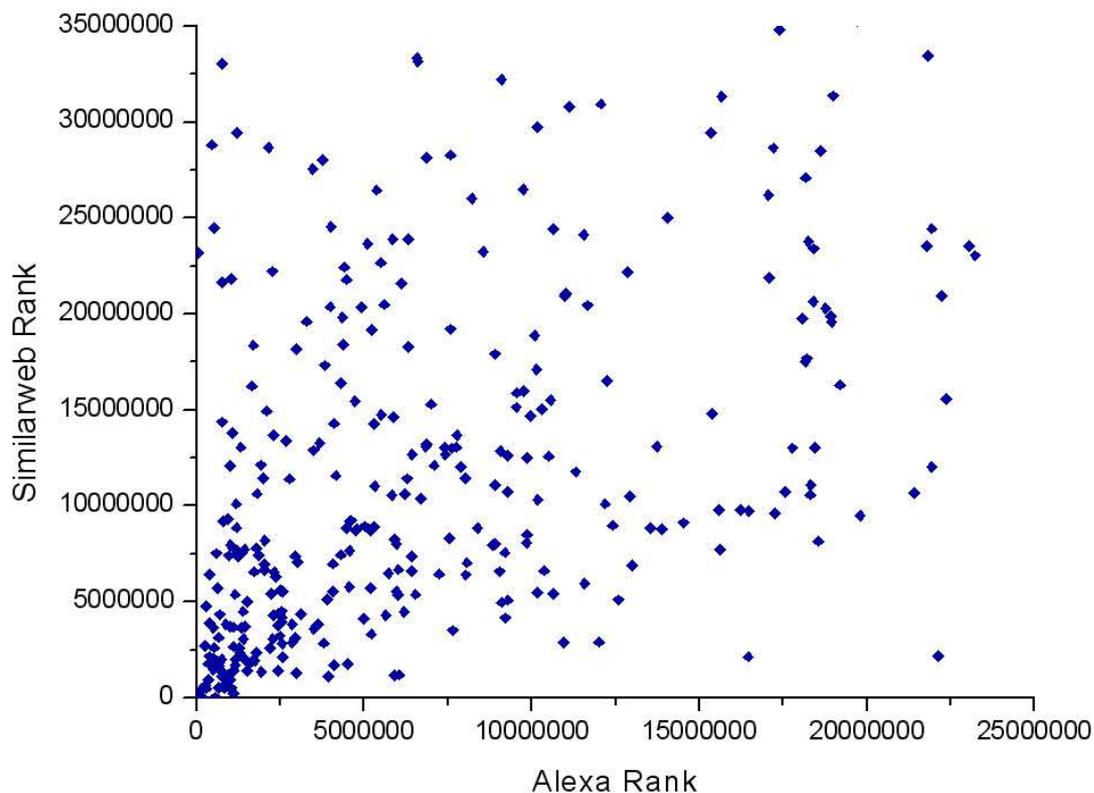

**Figure 2:** Relationship between SimilarWeb and Alexa ranks of the 903 investigated domains

## Domain Name Properties

**SEMrush search phrase volume (SV)**: Among other services, SEMrush collects and analyses Google search data, i.e., search volumes for keywords in the domain name; the rank of a domain for those keywords and the number of ads displayed in the search results of a given keyword.  The value of words in a domain name is characterised by its length and the keywords it contains. SEMrush search volume is proportional to the value of a keyword, thus high search volume domain names are worth more in most cases. While the topic of the keyword influences the value of the search volume as well, there is no way to programmatically determine the topic and the value of that topic from a given set of keywords.

**Total Extension Counts (TE)**: The principle behind this metric is that a more valuable domain name will be more likely to be registered under multiple popular TLDs. This value indicates the total number of currently registered domains with the same primary string under the extensions .com, .net, .org, .info, .biz, .us.

## Domain Health Properties

**Potential blocks (SB, PB, AB)**: Sites with inappropriate, malicious or duplicate content are routinely penalized by search engines, as are sites that are discovered to be spamming or flooding other sites such as unrelated forums and comment sections with backlinks to their own pages. The worst offenders are completely removed from Google's index – herein referred to as "Site Block" (SB) – and anecdotal evidence suggests that once de-indexed, it is extremely difficult, if not impossible to get back into the search giant's good graces. In this case a Google search using the site: prefix will yield no pages. Further, sites that serve inappropriate, adult or poor content, do not follow the guidelines prescribed by ad networks and/or engage in click fraud (clicking on ads on their own pages) can be penalized by being removed from the Google AdSense program (Kshetri 2010), the best known ad network for small and big web publishers alike. In our property descriptors this penalty is identified as "AdSense Block" (AB). This penalty is known to be easier to reverse by becoming compliant with AdSense rules. A third penalty, referred to here as "Park Block" (PB), signifies a state wherein Google refuses to serve ads on a parked domain. These assessable penalties, particularly site block, can have a negative impact on a site's ability to attract visitors or gain credibility via appearance in search results. The potentially poor existing backlink profile of such sites can also be a compounding factor.

In summary, tools quantifying domain properties were described above. The analysed aspects were domain name and domain traffic, backlink profile, history and domain health. Our analysis revealed that although some tools are aimed at analysing the same property, they possess low correlation thus indicating the inaccuracy of prediction methods. Consequently, no one property can be used as a universal measure; multiple properties should instead be utilized as parameters in a complex prediction model.

In the course of our study, 903 domain names auctioned at GoDaddy, NameJet and Snapnames were collected. All the domain properties described above were collected for each domain (**Supplementary Table 1**). It should be noted that **Table 1** is not a complete list of domain properties; other services such as ahrefs and MajesticSEO offer tools that have not been employed in our method. Furthermore, as discussed, there is overlap between various property sets (e.g., DP, DA, PA, BL and PR all analyse the backlinks of a domain, while Alexa rank and SimilarWeb rank both indicate current traffic).

The data in **Supplementary Table 1** was further analysed to test for the existence of any simple rules that might easily separate valuable domains from non-valuable ones; the results are summarized in **Table 2**.

|    | Sales price over 100 USD | | | | Sales price under 100 USD | | | |
|----|-----|-----|---------|--------|-----|-----|---------|--------|
|    | MIN | MAX | Average | median | MIN | MAX | Average | median |
| **PR** | 0 | 7 | 3.29 | 4 | 0 | 7 | 1.06 | 0 |
| **PA** | 1 | 72 | 42.23 | 46 | 1 | 56 | 16.51 | 18.5 |
| **DA** | 1 | 70 | 33.98 | 37 | 1 | 59 | 15.02 | 13 |
| **BL** | 0 | 322600 | 5369.22 | 331 | 0 | 33808 | 1376.34 | 0 |
| **DP** | 0 | 2741 | 278.70 | 156 | 0 | 2126 | 122.35 | 6 |
| **ACR** | 0 | 47256 | 396.27 | 201 | 0 | 19035 | 125.34 | 15 |
| **Alexa** | 24476 | 40000000 | 17062231 | 10249830 | 51330 | 40000000 | 33055530 | 40000000 |
| **Similarweb** | 28459 | 40000000 | 20912188 | 18731855 | 617889 | 40000000 | 35280489 | 40000000 |
| **DOB** | 1995 | 2016 | 2003 | 2002 | 1996 | 2016 | 2010 | 2011 |
| **PB** | 0 | 1 | 0.10 | 0 | 0 | 1 | 0.12 | 0 |
| **SB** | 0 | 1 | 0.12 | 0 | 0 | 1 | 0.25 | 0 |
| **AB** | 0 | 1 | 0.07 | 0 | 0 | 1 | 0.09 | 0 |
| **TE** | 0 | 6 | 2.87 | 2 | 1 | 6 | 2.07 | 1 |
| **SV** | 0 | 450000 | 3394 | 10 | 0 | 1000000 | 21435 | 0 |
| **LE** | 6 | 32 | 13.91 | 13 | 6 | 33 | 16 | 15 |

**Table 2: Analysis of domain data**

The most important finding of the analysis was that for all individual parameters the minimum and maximum values are not solely indicative of non-value or value. PR can range from 0 to 7 in both classes. As indicated before, PR has been a major target of manipulation by 'black hatters'; the high PR of non-valuable domains is in sharp contrast to their backlink profiles and thus can safely be dismissed as fake using other domain parameters e.g. DA, PA and DP. Both the average and median PR values are higher in the valuable group. Similarly, valuable domains tend to possess higher ACR, but since ACR can be increased manually, cases of isolated high ACR can be found in the non-valuable group as well. The year of registration was also not a simple predictor of domain value: 20-year old domains can be found in both groups. It bears repeating that the active age of a domain in our property set reflects the first time a page for the domain appears in the Wayback Machine archive, thus a DOB of 2016 can mean a newly registered domain or simply one that has never been captured by archive.org. Blocking parameters (PB, AB, and SB) in turn indicate that even blocked domains can be snapped up if other signals are strongly positive.
The final goal of our study was to develop a mathematical model using data from **Supplementary Table 1** that is able to identify valuable domains. Although it can be realistically assumed that domain prices in an open auction are close to a fair market, the sales prices still show large variations between each transaction of the same domain name. This variation can be traced to fluctuations in the domain market (Lindenthal 2016) or to a specific buyer's need for the specified domain name. The size of the variation can often be tenfold (as an example see: https://namebio.com/blog/daily-market-report-for-march-28th-2016/). Therefore, prediction of an exact domain value is limited and since the input data shows such a high diversity, the prediction of an exact sales price will be inaccurate. Moreover, the real life problem is not the determination of an exact price, but that potentially valuable domains should be discovered from hundreds of thousands of daily expiring or auctioned domains. Based on these two needs, classification of domains to valuable and non-valuable ones was our choice of method rather than exact price prediction. There are a number of statistical methods available aimed at solving binary classification problems such as Discriminant Analysis or SVM. The Support Vector Machines method has been widely and interdisciplinary used for binary classifications (Mjolsness 2001), (Noble 2006), (Tian, Shi, and Liu 2012).
In the SVM context domains can be presented in an N dimensional space according to N previously defined quantitative parameters. The SVM model is a representation of the domains as points in this space, so that the domains of the two categories (valuable and non-valuable ones) are divided by a separating surface. Domains not used in the course of SVM model development are then mapped into the same N dimensional space and predicted to belong to a category based on their location in this space. A key aspect of SVM is the optimization of the correct selection of descriptors (feature selection). Reducing the number of variables that describe a data point has several advantages, such as shorter training times, reduced probability of overfitting and easier interpretation of data. In our case instead of feature selection, feature extraction was carried out in order to offset the vulnerability of individual domain properties to data limitations and potential manipulation. For feature extraction the correlating descriptors can be defined and grouped with the means of hierarchical clustering (Park 2013).
**Figure 3** shows the results of hierarchical clustering of the investigated domain parameters. Clustering results were in accordance with the different aspects of domains.

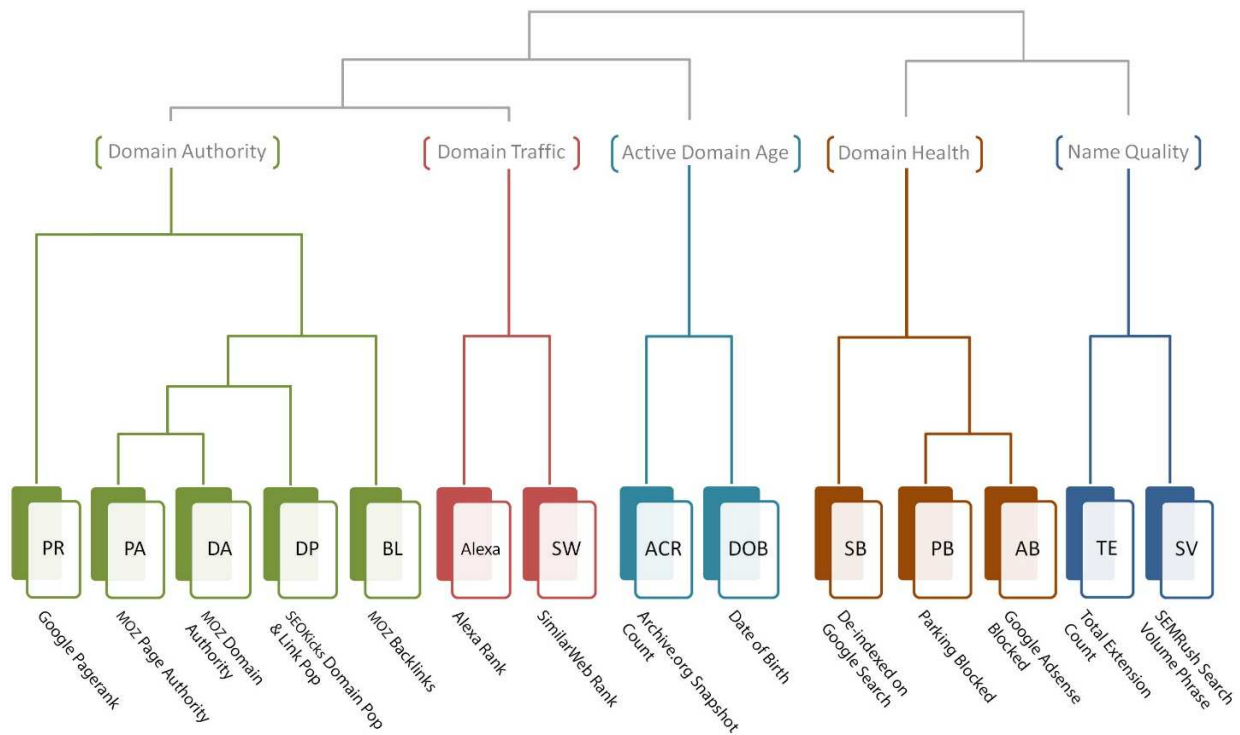

**Figure 3:** Hierarchical clustering of domain descriptors

The first cluster we call *Domain Authority* consists of PR, DA, PA, BL and DP. These descriptors are all calculated from the backlink profile of a domain, so their correlation is not surprising. DA and PA originate from the same web service (Moz) and thus possess the highest correlation. PR - being the most important descriptor in a domain's backlink profile - stands alone. This also reflects its significance on one hand and its vulnerability and anachronistic nature on the other. The significance of PR lies in the fact that Google is in the possession of all backlinks that are used for search engine ranking, while the other four properties are calculated by web crawlers that try to explore relevant links from a comparatively smaller data set.

The second cluster we call *Domain Traffic* consists of two parameters: SimilarWeb Rank and Alexa Rank. Both estimate the traffic of a domain using different methods. Thus their values correlate, especially for high traffic domains.

The third cluster we label *Active Domain Age* comprises the domain's registration year and the number of archive snapshots (ACR). Registration year obviously refers to the domain's age, whereas its archive snapshot indicates its usage history and gives us a rough estimation how popular it was.

The fourth cluster we call *Domain Health* contains the penalty parameters SB, AB and PB.

AdSense block and parking block further correlate as they respectively influence the ability of showing CPC ads on a domain with or without content.

The fifth cluster we categorize as *Name Quality* consists of the extension count and exact term search value for a domain. This indeed maps well to name quality: the better or more popular the name is, the higher the probability that more extensions are taken. Also, the higher the exact search term volume, the higher probability that the name reflects an important term. The most expensive domains in the valuable class indeed possess high exact search term volume and all extensions are taken.

These five clusters indicate five fundamental and quantifiable properties of domains. The individual clusters show low correlation with each other; thus they are well optimized for use as descriptors for our SVM study.

In chemistry, molecular descriptors are often calculated by the geometric mean of atomic properties e.g. Sanderson group electronegativity (Sanderson 1983) or WHIM symmetry (Todeschini and Gramatica 1997). Analogously, geometric means of domain properties belonging to the same cluster were used as descriptors. Geometric mean is a good summary statistical property as all properties are considered.

The five descriptors were scaled and SVM classification calculations were carried out in order to find the separating hyper-plane. The results are indicated in **Table 3**.

|  | TP | TN | FP | FN | ACC | SE | SP | MCC |
|---|---|---|---|---|---|---|---|---|
| training set | 152 | 158 | 2 | 10 | 0.963 | 0.938 | 0.988 | 0.927 |
| test set | 146 | 154 | 1 | 8 | 0.971 | 0.948 | 0.994 | 0.943 |
| external set | 120 | 118 | 11 | 7 | 0.930 | 0.945 | 0.915 | 0.860 |

**Table 3:** SVM results of the training. test and external sets. TP: true positive. TN: true negative. FP: false positive. FN: false negative. ACC: accuracy. SE: sensitivtiy. SP: specificity. MCC: Matthews correlation coefficient

As can be seen, our SVM model separated valuable domains from non-valuable ones with 97% accuracy in the test and 93% accuracy in the external data sets. It is important to emphasize that besides objective descriptors, there is a subjective factor in determining the value of a domain name. It can happen that the same domain name is sold once for more than and once for less than 100 USD. Therefore, reaching 100% accuracy is not possible and misclassification does not always reflect a problem with the model. Besides accuracy, the balance of the model is of great importance as well, namely, in terms of specificity (true negative rate i.e. is the model able to recognize all non-valuable domains?) and sensitivity (true positive rate i.e. is the model able to recognize all valuable domains?). Also, Matthews's correlation coefficient was calculated in order to evaluate the correlation between observed and calculated classifications. In our study we aimed at a balanced SVM model i.e. with similar specificity and sensitivity. The MCC was above 85% in case of all three datasets.

Further analysis of the false negative results (i.e. where the real auction price was above 100 USD, whereas the predicted price was lower) reveals that the most expensive falsely predicted domains fall into a so-called brandable domain category: dataclick.com, finerugs.com and golfportal.com. In these cases the auction value was based on the catchiness of the name and its potential to support or create a brand such as a blog or business. This quality of the name was not reflected in the exact search volume; furthermore, the domains were relatively young and few other extensions were registered with the same name. In the case of finerugs.com, all other extensions were available for registration and after purchase the domain was redirected to finerugs.co.za, indicating that the final price was determined by factors other than the inherent value of the domain name. Analysing false positive domains that were actually unsold (i.e. where our SVM model predicted a valuable domain) reveals a possible manipulation (i.e. either there is a substantial difference between Alexa and SimilarWeb ranks or between DA and PA values).

**Practical use and further plans**

Our SVM model can be used to routinely screen aftermarket domains for the automatic identification of valuable ones. According to expireddomains.net, there are currently over 869,000 domains in 'pending delete' status on various registrars. There are another 368,600 domains on GoDaddy's expired domains list, 284,000 in NameJet's pre-release and 418,000 in Snapnames' pre-release phases. Running our SVM model identified 550 valuable domains among these. While manual investigation of all available domains is impossible, the filtered list can be easily investigated by domain investors and can be further refined by actual needs or personal interests. The current balanced model can also be attuned to increase either sensitivity or specificity. A model with 100% sensitivity would ensure that no valuable domain will be lost during the screen but the list would contain more false positives. Developing a model with 100% specificity would ensure that all domains predicted to be valuable are real positives, so that backordering of all domains can be automated for large volume domain investors. Further refinement of the model could be carried out by developing a method for identification of brandable domain names.

# Conclusion

In the course of our study various parameters correlating to domain aftermarket valuations were identified. The domain aftermarket is a billion dollar industry and domain names can be characterized by various metrics that are not randomly distributed. These metrics were extracted from multiple tools that investigate the quality and relevance of words in the domain name and the history of the domain. Due to incomplete data and potential manipulations, correlating parameters were taken into account for the determination of domain value. Analysing available parameters led to the determination of five composing factors of domain value: the name, authority, traffic, active age and health of the domain. A high throughput SVM method was developed based on the descriptors that was able to differentiate between valuable and non-valuable domains with 95% accuracy and can be used to effectively identify valuable domains in the deluge of the domain aftermarket.

# Acknowledgement

The authors are grateful to Siddesh Mukundan, Kanak Arora and authors of domain.tips for fruitful discussions and help in preparation of this manuscript.